\documentclass[prm,twocolumn, amsmath,amssymb,
reprint,%
author-year,%
author-numerical,%
dvipdfmx,
]{revtex4-1}

\usepackage{graphicx}
\usepackage{bm}
\usepackage{color}

\begin{document}

\title{Site preference of chalcogen atoms in 1T$^\prime$ $MX_{2(1-x)}Y_{2x}$ ($M=$ Mo and W; $X, Y=$ S, Se, and Te)}
\author{Shota Ono}
\email{shotaono@muroran-it.ac.jp}
\author{Ryotaro Ohse}
\affiliation{Department of Sciences and Informatics, Muroran Institute of Technology, Muroran 050-8585, Japan}

\begin{abstract}
The insulator-metal transition, accompanying the structural phase transition from 2H to 1T$^\prime$ structure, has been reported in two-dimensional W-S-Te and W-Se-Te systems. It is also reported that Te atoms tend to occupy a specific site of the 1T$^\prime$ structure. Here, we study the site preference of chalcogen atoms in $MX_{2(1-x)}Y_{2x}$ ($M=$ Mo and W; $X, Y=$ S, Se, and Te; $0\le x \le 1$) using first-principles approach. We demonstrate that the site preference of chalcogen atoms explains the universal correlation between the formation energy and the Peierls-like distortion amplitude in the 1T$^\prime$ phase. The impact of the site preference on the linear elastic properties is strong, whereas its impact is weak in the non-linear regime. This establishes the structure-property relationships in $MX_{2(1-x)}Y_{2x}$ systems. 
\end{abstract}

\maketitle

\section{Introduction}
Among many two-dimensional (2D) materials, transition metal dichalcogenides $MX_2$ ($M=$ Mo and W; $X=$ S, Se, and Te) have been widely studied due to their phase-dependent properties. The monolayers except for WTe$_2$ adopt the 2H structure, where $X$, $M$, and $X$ hexagonal layers form the ABA stacking. They have a direct bandgap of a few eV at K and K$^\prime$ points in the Brillouin zone and serve as candidate materials for quantum technologies. On the other hand, WTe$_2$ monolayer adopts the 1T$^\prime$ structure, where $X$, $M$, and $X$ hexagonal layers form the ABC stacking (1T-type) and a Peierls-like lattice distortion is present \cite{zhao2021}. This system serves as a 2D topological insulator \cite{qian2014} and a ferroelectric and shape memory material \cite{li2016}. The phase transformation between 2H and 1T$^\prime$ structures for $MX_2$ has been investigated using first-principles calculations \cite{duerloo2014}, which paves the way to the phase engineering of 2D materials and their applications \cite{li2021,chen2022,kim2024}.



The 2D alloys based on $M$-$X$-$Y$ systems have been created experimentally \cite{yao2020,yun2017,zhao2019,duan2016,ashkan2021,wang2020,yu2017,oliver2020,suenaga2018}. For example, WS$_{2(1-x)}$Se$_{2x}$ alloys was created by using colloidal growth method \cite{ashkan2021}; WS$_{2(1-x)}$Te$_{2x}$ was synthesized by chemical vapor deposition \cite{wang2020}; and WSe$_{2(1-x)}$Te$_{2x}$ bulk crystals was grown by chemical vapor transport and the mechanical exfoliation method was used to produce their monolayers \cite{yu2017,oliver2020}. WS$_2$ and WSe$_2$ have finite bandgap, while WTe$_2$ is metallic. Therefore, an insulator-metal transition (IMT) should occur with increasing Te concentration. The IMT has been observed at $x\simeq 0.5$ in WS$_{2(1-x)}$Te$_{2x}$ \cite{wang2020} and WSe$_{2(1-x)}$Te$_{2x}$ \cite{yu2017,oliver2020} through optical bandgap measurements. Simultaneously, the phase transition from 2H to 1T$^\prime$ has been confirmed through Raman spectroscopy and direct observations. Thermodynamics of 2D $MX_{2(1-x)}Y_{2x}$ alloys has been studied by using first-principles approach, and the mixing entropy is suggested to be an important role in stabilizing the alloy phase \cite{komsa2012,silva2022}. However, a nontrivial site preference of chalcogen atoms has been reported in W$X_{2(1-x)}$Te$_{2x}$ ($X=$ S and Se) \cite{wang2020,suenaga2018}, where $X$ and Te atoms tend to occupy different region in the unit cell. This indicates that $M$-$X$-$Y$ systems in the 1T$^\prime$ phase are no longer ideal alloys. Therefore, it is important to understand which specific configurations stabilize the 1T$^\prime$ phase.

Due to the site preference of Te atoms \cite{wang2020,suenaga2018}, the local bonding environment becomes anisotropic. The in-plane elastic constants reflect the energy change with respect to lattice strain and the relaxation of the atomic sites, i.e., the curvature of the potential energy surface around the equilibrium geometry. Such curvature may be similar for the 1T$^\prime$ structure with similar site occupations, and therefore the linear elastic constants are expected to be largely determined by the site preference. 

In this paper, we investigate the site preference of chalcogen atoms in $MX_{2(1-x)}Y_{2x}$ ($0\le x \le 1$) systems and predict a possible impact on the elastic properties using first-principles approach. By studying the energetic stability of various configurations of chalcogen atoms $X$ and $Y$ in the 1T$^\prime$ structure, we show that a universal correlation exists between the Peierls-like distortion and formation energy in these systems, and demonstrate that this trend reflects the site preference of $Y$ atoms in the elongated region in the unit cell. We also show that the site preference of Te atoms has a strong impact on the in-plane elastic properties in the linear regime, whereas its impact is weak in the non-linear regime. The present work establishes the structure-property relationships in $MX_{2(1-x)}Y_{2x}$ ($0\le x \le 1$) systems. 

\begin{figure*}
\center\includegraphics[scale=0.55]{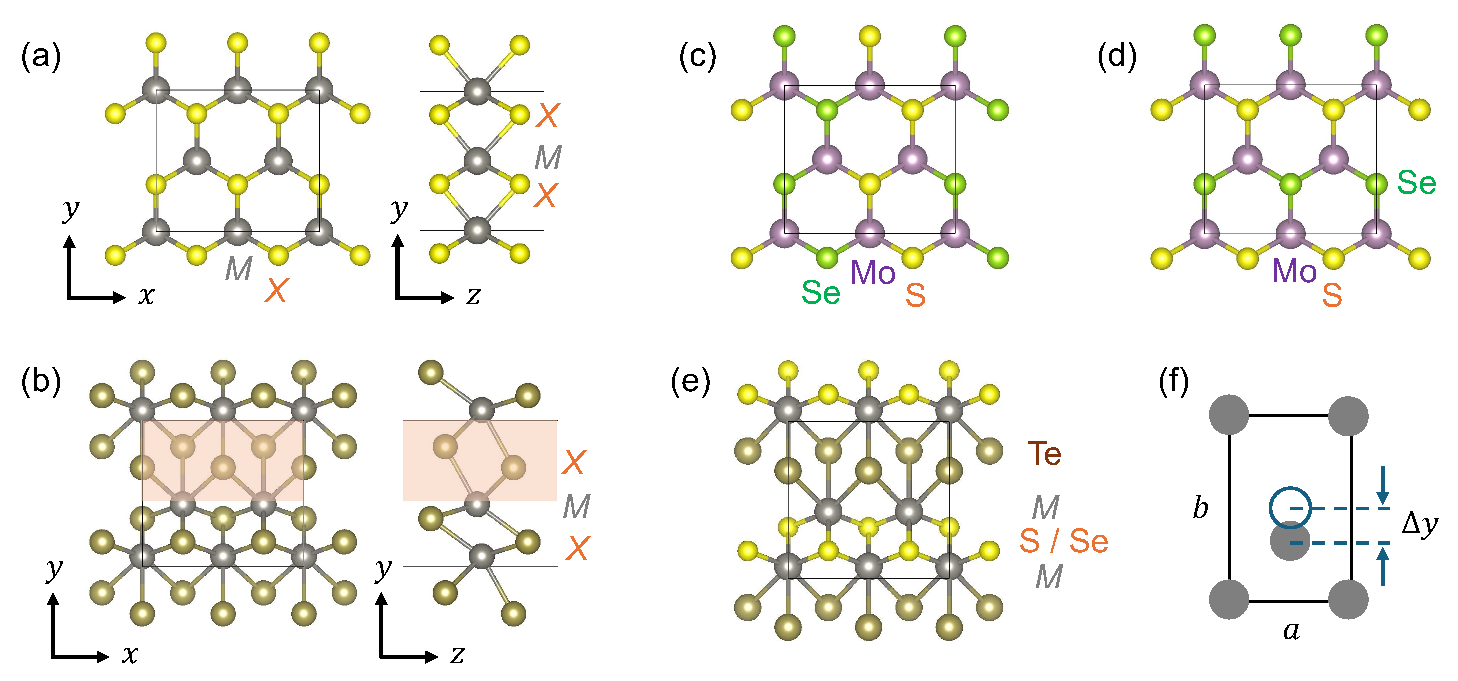}
\caption{Top and side views of (a) 2H and (b) 1T$^\prime$ $MX_2$ in a $2\times 1$ supercell. The metal atom $M$ shifts to $-y$ direction (a Peierls-like distortion) in the 1T$^\prime$ phase, which separates into the compressed and elongated (shaded) region. Top view of (c) the lowest and (d) the second lowest energy structure of 2H $M$SSe [$M=$ Mo and Te (purple)]. The chalcogen atoms in the bottom side have different atomic species of the upper side [i.e., S $\leftrightarrow$ Se (yellow and green)]. (e) The lowest energy structure of 1T$^\prime$ Mo$X_{2(1-x)}$Te$_{{2x}}$ and W$X_{2(1-x)}$Te$_{{2x}}$ at $x=0.5$. Te atoms are located in the elongated region [i.e., the shaded area in (b)]. (f) Schematic illustration of the Peierls-like distortion, where $M$ atom in the center exhibits a displacement of $\Delta y$. } \label{fig1} 
\end{figure*}

\section{Computational details}
We considered 2H and 1T$^\prime$ structures and calculated the total energies of $MX_{2(1-x)}Y_{2x}$ ($0\le x \le 1$) monolayer assuming a $2\times 1$ supercell of the rectangular unit cell (see Fig.~\ref{fig1}(a) and (b)). There are 12 atoms in the supercell, and the number of configurations is 34 and 55 for 2H and 1T$^\prime$ structures, respectively, where the inequivalent structures were enumerated using StructureMatcher \cite{pymatgen}. The total energy of these structures were calculated using Quantum ESPRESSO \cite{qe}. The generalized-gradient approximation (GGA) parametrized by Perdew, Burke, and Ernzerhof (PBE) \cite{pbe} was used for the exchange-correlation energy functional. The pslibrary \cite{pslibrary} was used for the pseudopotentials of W, Te, and S atoms (Mo.pbe-spn-, W.pbe-spfn-, S.pbe-nl-, Se.pbe-n-, and Te.pbe-n-rrkjus\_psl.1.0.0.UPF). Spin-polarized calculations were performed. The energy cutoff for wavefunctions and charge density was set to be 60 and 480 Ry, respectively. Convergence thresholds for the total energy in the self-consistent field calculations was set to be $10^{-6}$ Ry, and those for the total energy and forces for structure optimization were set to be $10^{-4}$ Ry and $10^{-3}$ a.u., respectively. We assumed the smearing parameter of 0.01 Ry \cite{smearingMV} and $k$-point distance $\Delta k$ smaller than 0.15 \AA$^{-1}$. The orthorhombic unit cell was used and the length of the $c$ axis was fixed to 20 \AA \ to avoid the spurious interaction between slabs along the out-of-plane direction. Crystal structures were visualized using VESTA \cite{vesta}.

In the present work, we define the formation energy as
\begin{eqnarray}
 \Delta E_\gamma (x) = E_\gamma (x) - (1-x)E(MX_2) - xE(MY_2),
 \label{eq:formation_energy}
\end{eqnarray}
where $E_\gamma (x)$ is the total energy per atom for $MX_{2(1-x)}Y_{2x}$ monolayer with the structure $\gamma$ and $E(MX_2)$ is the total energy per atom of $MX_2$ monolayer in the ground state structure. $\gamma$ specifies the prototype of 2H or 1T$^\prime$ and the spacial distribution ($\alpha$) of $X$ and $Y$ atoms. 

2H $MX_{2(1-x)}Y_{2x}$ in the 1T$^\prime$ structure should include large strain energy that originates from the 2H-1T$^\prime$ phase transition accompanying the structural relaxation. To study the stability of the 1T$^\prime$ structure with a configuration $\alpha$, we decompose the formation energy into the phase transition energy $\Delta E_{\rm PT}$ and the relaxation energy $\Delta E_{\rm relax}$ defined as
\begin{eqnarray}
 \Delta E_{{\rm 1T^\prime},\alpha} (x) &=& \Delta E_{\rm PT}(x) + \Delta E_{\rm relax}(x),  \\
 \Delta E_{\rm PT}(x) &=& x \left[ E_{\rm 1T^\prime} (MX_2) - E (MX_2) \right] \nonumber\\
 &+& (1-x) \left[ E_{\rm 1T^\prime} (MY_2) - E (MY_2) \right], \\
 \Delta E_{\rm relax}(x) &=& E_{{\rm 1T^\prime},\alpha} (x) -  x E_{\rm 1T^\prime} (MX_2) \nonumber\\
 &-& (1-x) E_{\rm 1T^\prime} (MY_2). 
\end{eqnarray}
$E_{\rm 1T^\prime} (MX_2)$ is the total energy of 1T$^\prime$ $MX_2$, and therefore, the contribution from WTe$_2$ to $\Delta E_{\rm PT}$ vanishes. 


The 2D material stores the strain energy $E_{\rm s}$ when the nano-sheet is elongated along the in-plane directions. In the linear regime, $E_{\rm s}$ is expanded up to second order in strain $\varepsilon_i \ (i,j = 1,2)$, where $i=1$ and 2 indicates the direction of $x$ and $y$, respectively. The elastic constants are defined by 
\begin{eqnarray}
 c_{ij} = \frac{1}{S(0)}\frac{\partial^2 E_{\rm s}}{\partial \varepsilon_{i}\partial \varepsilon_{j}},
 \label{eq:cij_def}
\end{eqnarray}
where the derivative is calculated at the relaxed geometry with the surface area of $S(0)$. In the nonlinear regime, $E_{\rm s}$ no longer exhibits a quadratic dependence on $\varepsilon_i$. The stress at a strain $\varepsilon_i$ is defined by
\begin{eqnarray}
 \sigma_i = \frac{1}{S(\varepsilon_i)}\frac{\partial E_{\rm s}}{\partial \varepsilon_i},
\end{eqnarray}
where $S(\varepsilon_i)$ is the surface area at $\varepsilon_i$, assuming the uniaxial strain. In the present work, the derivatives are approximated using the finite-difference method. 


\begin{figure*}
\center\includegraphics[scale=0.7]{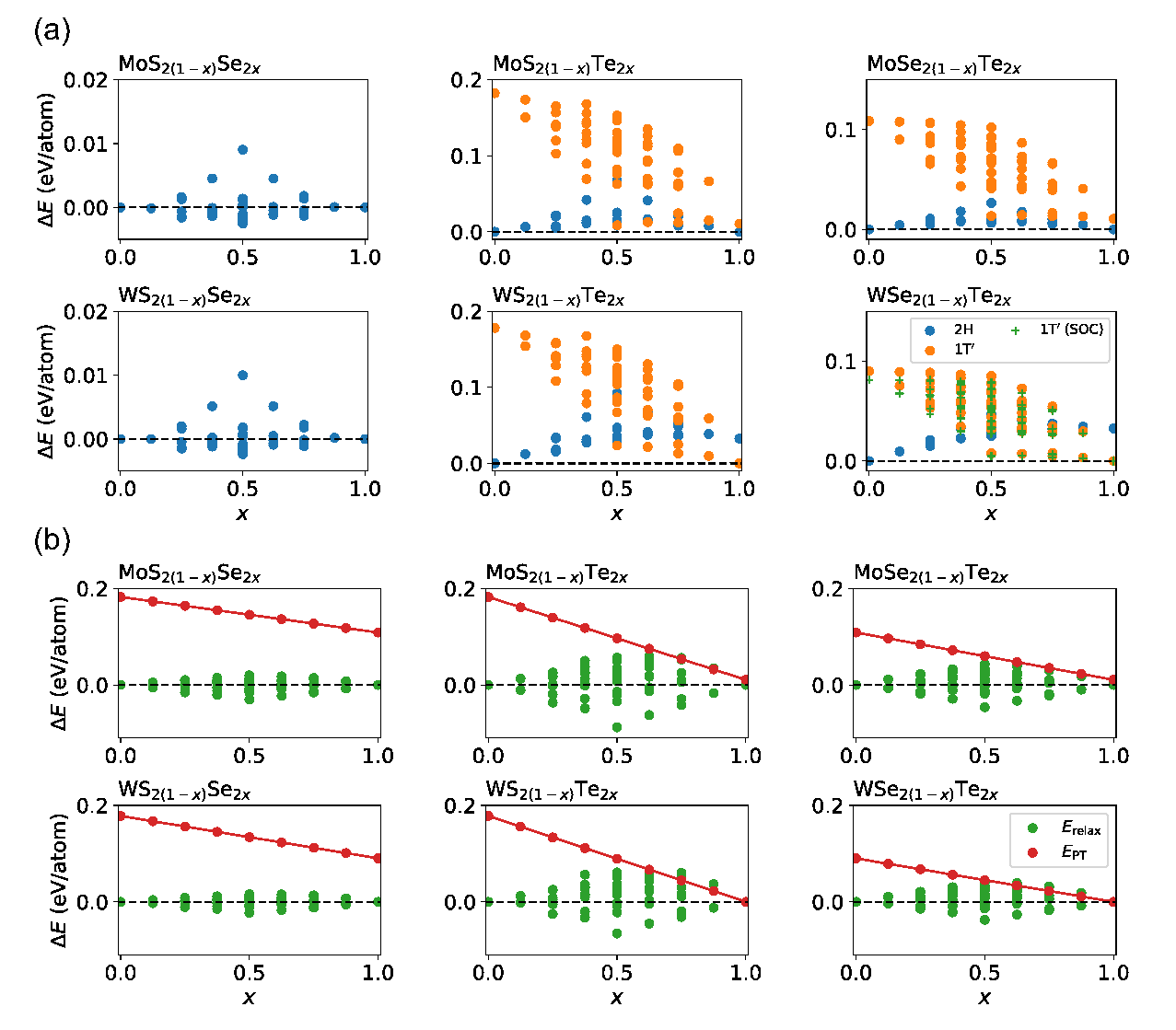}
\caption{(a) $\Delta E_\gamma$ for $MX_{2(1-x)}Y_{2x}$ alloys with 2H and 1T$^\prime$ structures. $M$S$_{2(1-x)}$Se$_{2x}$ alloys ($M=$ Mo and W) in the 1T$^\prime$ structure has $\Delta E > 0.1$ eV/atom that is outside of the energy window. $\Delta E$ including the SOC is also plotted for 1T$^\prime$ WSe$_{2(1-x)}$Te$_{2x}$. (b) $\Delta E_{\rm PT}$ and $\Delta E_{\rm relax}$ for 1T$^\prime$ structure. } \label{fig2} 
\end{figure*}

\begin{figure*}
\center\includegraphics[scale=0.5]{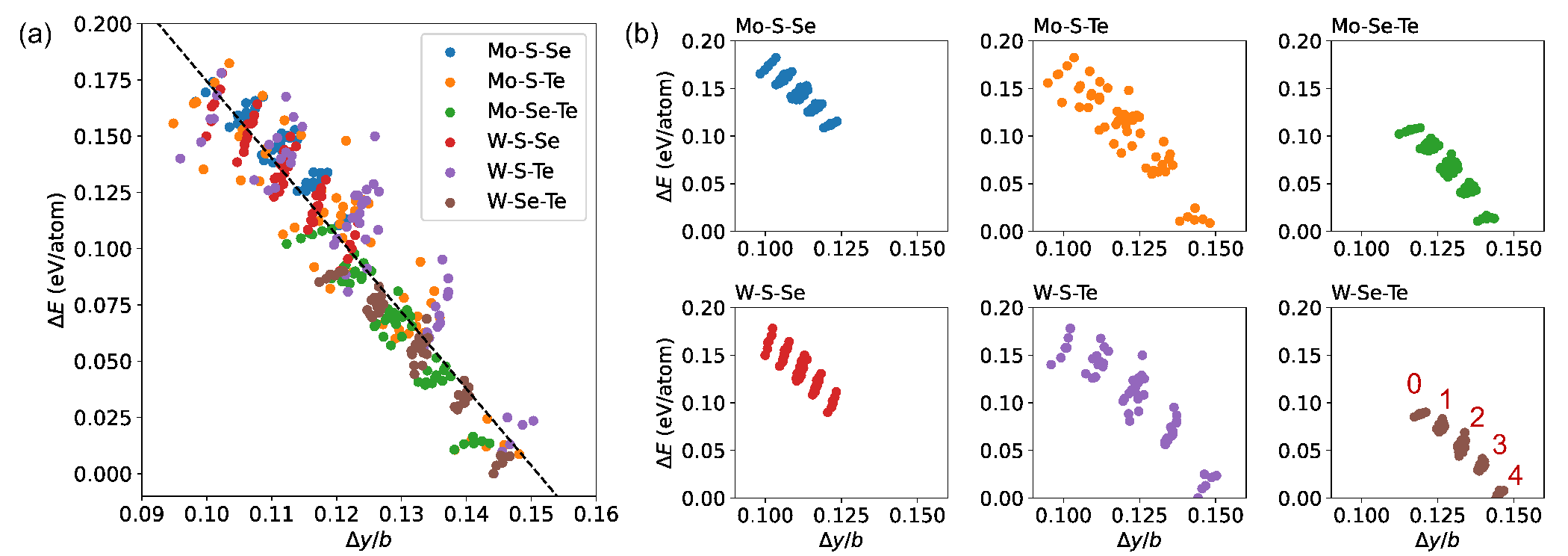}
\caption{(a) $\Delta E$ as a function of the $M$ atom displacement in the 1T$^\prime$ structure relative to the 1T structure. The dashed line is a linear fit to the data ($\Delta E = -3.418\Delta y/b + 0.516$). (b) $\Delta E$ versus $\Delta y/b$ for each $M$-$X$-$Y$. The data of $(\Delta y/b, \Delta E)$ is classed into five groups indicated as $m \ (=0, 1, 2, 3, 4)$. } \label{fig3} 
\end{figure*}

\section{Results and Discussion}
\subsection{Structural properties}
Figure \ref{fig2}(a) shows the $x$-dependence of $\Delta E$ defined by Eq.~(\ref{eq:formation_energy}) for $MX_{2(1-x)}Y_{2x}$. For MoS$_{2(1-x)}$Se$_{2x}$ and WS$_{2(1-x)}$Se$_{2x}$, we observe $\Delta E <0$, indicating that mixing between S and Se atoms is preferable. $\Delta E$ has a minimum value at $x=0.5$. Figures \ref{fig1}(c) and \ref{fig1}(d) show the lowest and the second lowest energy structures, respectively. S and Se atoms are located along the armchair [Fig.~\ref{fig1}(c)] and zigzag [Fig.~\ref{fig1}(d)] chains, and the different atomic species appear on the bottom side of the 2H monolayer. Note that Janus-type $M$SSe, where one side of the $M$S$_2$ surface is fully replaced with Se, has positive $\Delta E$ (0.009 eV/atom and 0.01 eV/atom for $M=$Mo and W, respectively), while the Janus structure has been realized experimentally \cite{lu2017,zhang2017,lin2020}. For 1T$^\prime$ structure, $\Delta E$ is positive and higher than the 2H structure by an order of magnitude.  

For the other systems with $Y=$ Te, $\Delta E$ is positive for all $x$ and $\alpha$. The $\Delta E$ of the 1T$^\prime$ structure tends to be larger than that of 2H structure. Interestingly, one can find relatively stable structures for $x\ge 0.5$, and these 1T$^\prime$ structures become more stable than the 2H structures for WS$_{2(1-x)}$Te$_{2x}$ and WSe$_{2(1-x)}$Te$_{2x}$. 

The effect of spin-orbit coupling (SOC) was investigated for 1T$^\prime$ WSe$_{2(1-x)}$Te$_{2x}$. $\Delta E$ decreases for all $\alpha$, and the lowest $\Delta E$ for $x\ge 0.5$ is less than 0.005 eV/atom.  

$\Delta E_{\rm PT}$ and $\Delta E_{\rm relax}$ are shown in Fig.~\ref{fig2}(b). $\Delta E_{\rm PT}$ decreases with $x$, indicating that the energy cost for taking the 1T$^\prime$ structure decreases with increasing the Te concentration. Given the formation of the 1T$^\prime$ structure, $\Delta E_{\rm relax}$ has negative value and takes a minimum at $x=0.5$. This is consistent with previous experiments \cite{wang2020,yu2017,oliver2020,suenaga2018}, where alloying between $MX_2$ and $MY_2$ are observed. 

It should be noted that $\Delta E_{\rm relax}$ takes negative value also for MoS$_{2(1-x)}$Se$_{2x}$ and WS$_{2(1-x)}$Se$_{2x}$. This suggests that alloying is energetically preferable if the 1T$^\prime$ phase is assumed. In fact, the 1T$^\prime$ WS$_{2(1-x)}$Se$_{2x}$ has been synthesized experimentally, whereas such a phase is transformed into 2H phase by annealing at 380$^\circ$C \cite{ashkan2021}.

\begin{figure*}
\center\includegraphics[scale=0.45]{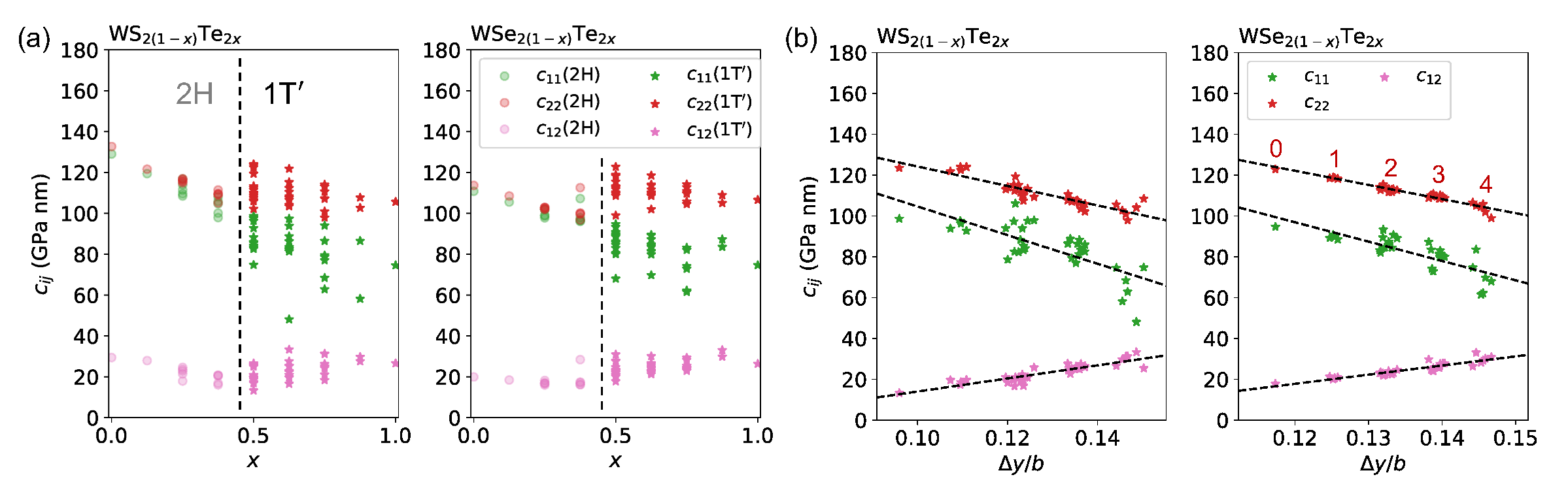}
\caption{(a) Distribution of elastic constants $c_{ij}$ in W$X_{2(1-x)}$Te$_{2x}$ ($X=$ S and Se). $c_{ij}$ of 2H and 1T$^\prime$ is plotted for $x < 0.5$ and $x\ge 0.5$, respectively. (b) The $\Delta y/b$-dependence of $c_{ij}$ in the 1T$^\prime$ phase with $x\ge 0.5$. The dashed line is a linear fit to the data. } \label{fig4} 
\end{figure*}

Having established the phase transition at $x=0.5$ in W-S-Te and W-Se-Te systems, we study the structural properties of 1T$^\prime$ phase. The 1T$^\prime$ structure is considered as a distorted phase of 1T structure, where the $M$ atoms are displaced from $y = b/2$ to $b/2 - \Delta y$ [see Fig.~\ref{fig1}(f)]. Figure \ref{fig3}(a) shows $\Delta E$ as a function of $\Delta y/b$ for 1T$^\prime$ $MX_{2(1-x)}Y_{2x}$. When $\Delta y/b$ is different between neighbor cells in the $2\times 1$ supercell, the averaged value was assumed. Irrespective to atomic species, the calculated data are well fitted by a liner curve of
\begin{eqnarray}
\Delta E = -3.418\frac{\Delta y}{b} + 0.516 \ \ {\rm (eV/atom)}.
\end{eqnarray}
$M$-S-Se systems have small distortion and large $\Delta E$; $M$-S-Te systems have broad range of $\Delta y/b$ and $\Delta E$; and $M$-Se-Te systems have large distortion and small $\Delta E$. 

Interestingly, the $(\Delta y/b, \Delta E)$ data for each $M$-$X$-$Y$ system fall into five groups [see Fig.~\ref{fig3}(b) for details]. The classification of such groups in Fig.~\ref{fig3} reflects the number of $Y$ (heavier) atoms in the elongated region [see shaded area with $y\in (0.5-\Delta y, b)$ in Fig.~\ref{fig1}(b)]. For the $2\times 1$ supercell, there are four sites that should be occupied by chalcogen atoms, and the number of occupation for $Y$ atoms can take five integers from $m=0$ to 4. As $m$ increases, $\Delta E$ decreases, yielding five groups. The lowest energy 1T$^\prime$ structure at $x=0.5$ is shown in Fig.~\ref{fig1}(e), which corresponds to $m=4$. For an $N\times 1$ supercell, the ($\Delta E,\Delta y/b$) data will fall into $2N+1$ groups because $2N$ sites for the chalcogen atoms exist in the elongated region. In this way, the number of Te atoms in the elongated region (i.e., the site preference) determines the amplitude of the Peierls-like distortion and the formation energy. 

Within the same group, $\Delta E$ tends to be large if the number of lighter atoms increases. This reflects the overall trend in Fig.~\ref{fig2}(a). 

\begin{figure*}
\center\includegraphics[scale=0.42]{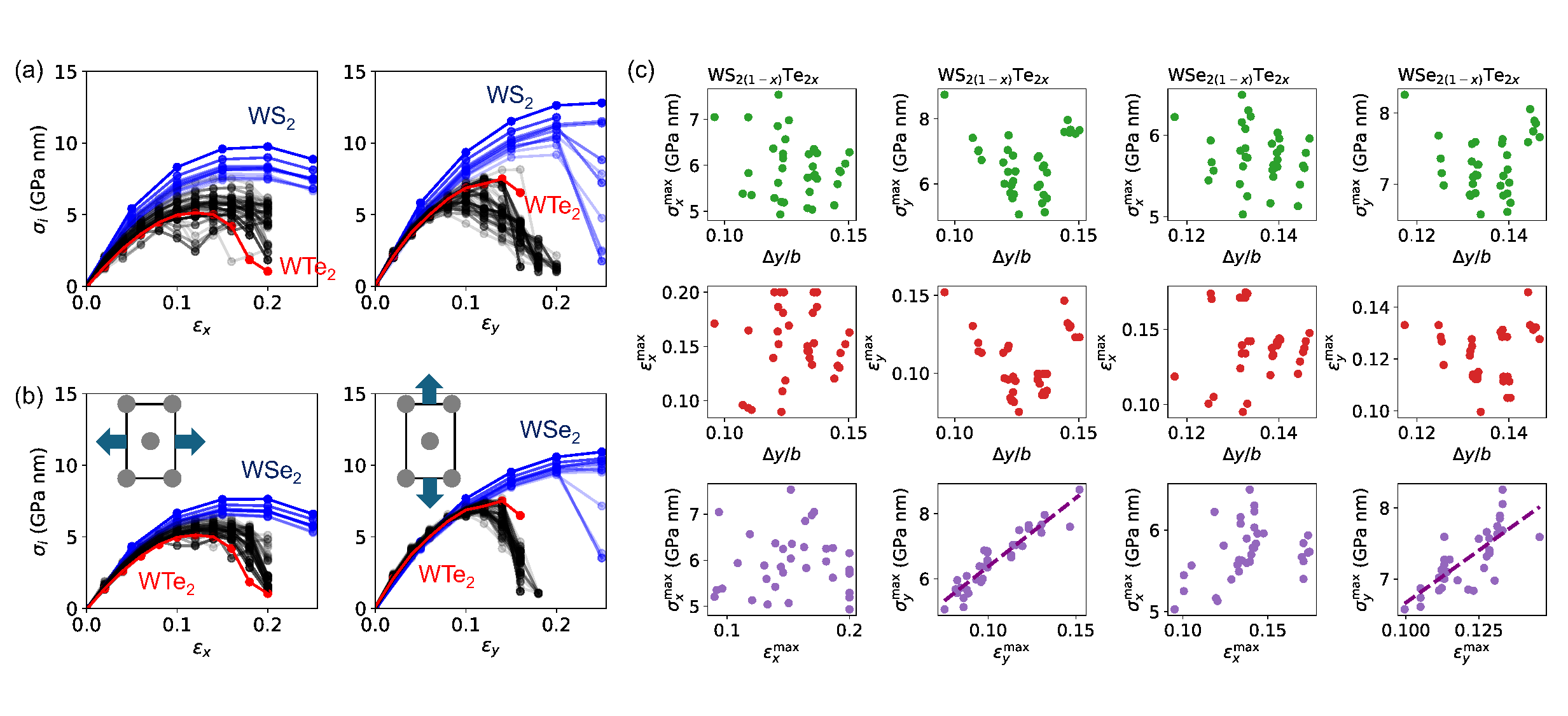}
\caption{Stress-strain curve for (a) WS$_{2(1-x)}$Te$_{2x}$ and (b) WS$_{2(1-x)}$Te$_{2x}$ with respect to the tensile strain along $x$ and $y$ directions. 2H (blue) and 1T$^\prime$ (black) phases were assumed for $x<0.5$ and $x\ge 0.5$, respectively. The curves become partially transparent as $x$ approaches 0.5. The curves for WTe$_2$ (i.e., $x=1$) are colored red. (c) The distribution of the maximum stress $\sigma_{i}^{\rm max}$, the maximum strain $\varepsilon_{i}^{\rm max}$, and the normalized displacement $\Delta y/b$ for WS$_{2(1-x)}$Te$_{2x}$ and WSe$_{2(1-x)}$Te$_{2x}$ in the 1T$^\prime$ phase. The dashed line is a linear fit to the data. } \label{fig5} 
\end{figure*}

\begin{table}\begin{center}\caption{The fitting parameters of $A$ and $B$ (in units of GPa nm) in Eq.~(\ref{eq:cij}). }
{\begin{tabular}{lcrr}\hline\hline
$M$-$X$-$Y$ \hspace{5mm} & $c_{ij}$ \hspace{5mm} & $A$ \hspace{5mm} & $B$ \hspace{0mm}  \\
\hline
W-S-Te \hspace{5mm} & $c_{11} $\hspace{5mm} & $-700.6$ \hspace{5mm} & $174.7$ \hspace{0mm}  \\
W-S-Te \hspace{5mm} & $c_{22} $\hspace{5mm} & $-476.6$ \hspace{5mm} & $171.9$ \hspace{0mm}  \\
W-S-Te \hspace{5mm} & $c_{12} $\hspace{5mm} & $320.4$ \hspace{5mm} & $-18.0$ \hspace{0mm}  \\
W-Se-Te \hspace{5mm} & $c_{11} $\hspace{5mm} & $-947.62$ \hspace{5mm} & 210.6 \hspace{0mm}  \\
W-Se-Te \hspace{5mm} & $c_{22} $\hspace{5mm} & $-692.6$ \hspace{5mm} & 205.2 \hspace{0mm}  \\
W-Se-Te \hspace{5mm} & $c_{12} $\hspace{5mm} & $448.0$ \hspace{5mm} & $-36.0$ \hspace{0mm}  \\
\hline
\end{tabular}}\label{table1}\end{center} \end{table}

\subsection{Linear and non-linear elastic properties}
Figure \ref{fig4}(a) shows the in-plane elastic constants $c_{ij}$ for W$X_{2(1-x)}$Te$_{2x}$, where the structure is switched from 2H to 1T$^\prime$ at $x=0.5$. Before the 2H-1T$^\prime$ phase transition, $c_{11}\simeq c_{22}$ and these are larger than $c_{12}$. In addition, $c_{ij}$ tends to decrease with increasing $x$, indicating the in-plane softening by alloying. It is interesting that the configuration-dependence ($\alpha$) is weak. On the other hand, when $x\ge 0.5$, $c_{22}$ becomes larger than $c_{11}$, and the values of $c_{ij}$ are scattered. However, $c_{11}$ and $c_{22}$ are relatively small for the lowest energy structure at each $x\in [0.5,1)$ (e.g., WS$_{2(1-x)}$Te$_{2x}$ at $x=0.625$ has $c_{11}=48$ GPa nm). If these structures dominate the area of W$X_{2(1-x)}$Te$_{2x}$ sheet, anomalous decrease in the elastic constants will be observed across $x=0.5$. 

To study the structure-property relationship in detail, we fitted the calculated data for 1T$^\prime$ phase by assuming a linear relation
\begin{eqnarray}
c_{ij} = A \frac{\Delta y}{b} + B  \ \ {\rm (GPa \ nm)}.
\label{eq:cij}
\end{eqnarray}
As shown in Fig.~\ref{fig4}(b), the data points, $(\Delta y/b, c_{ij})$, are well described by Eq.~(\ref{eq:cij}), and those fall into five groups again. This clearly shows that site preference determines the elastic constants in 1T$^\prime$ phase, while a deviation from the linear fit is larger in WS$_{2(1-x)}$Te$_{2x}$. The optimized parameters are listed in Table \ref{table1}. 

Figure \ref{fig5}(a-b) shows the stress-strain curve for WS$_{2(1-x)}$Te$_{2x}$ and WSe$_{2(1-x)}$Te$_{2x}$. The uniaxial strain along the direction $i \ (=x, y)$ was assumed. 2H WS$_2$ and WSe$_2$ (i.e., $x=0$) shows the largest values of the maximum stress $\sigma_{i}^{\rm max}$ and the maximum strain $\varepsilon_{i}^{\rm max}$ for both directions $i$. The non-linear elastic properties are anisotropic: For 2H WS$_2$, $\sigma_{x}^{\rm max}=9.7$ GPa nm at $\varepsilon_{x}^{\rm max} = 0.2$, while $\sigma_{y}^{\rm max}=12.8$ GPa nm at $\varepsilon_{y}^{\rm max} = 0.25$. The value of $\sigma_{i}^{\rm max}$ of 2H WSe$_2$ is smaller than that of 2H WS$_2$. As Te concentration increases, the stress-strain curve exhibits a downward shift. When the 1T$^\prime$ phase is formed (i.e., $x\ge 0.5$), both $\sigma_{i}^{\rm max}$ and $\varepsilon_{i}^{\rm max}$ exhibit significant decrease. A pronounced drop of $\sigma_i$ is observed for $\varepsilon_i \gtrsim 0.1$, and at large $\varepsilon_y$, the nanosheet splits into two parts. 

Figure \ref{fig5}(c) shows the $\Delta y/b$-dependence of $\sigma_{i}^{\rm max}$ and $\varepsilon_{i}^{\rm max}$. The values of $\sigma_{i}^{\rm max}$ and $\varepsilon_{i}^{\rm max}$ are scattered for each group characterized by $\Delta y/b$, especially for group of $m=1, 2$, and 3. This indicates that the non-linear elastic properties are not solely determined by the site preference. However, a linear correlation between $\sigma_{y}^{\rm max}$ and $\varepsilon_{y}^{\rm max}$ is observed, whereas the correlation is weak for WSe$_{2(1-x)}$Te$_{2x}$ [see the bottom in Fig.~\ref{fig5}(c)]. 

Note that anisotropy of the non-linear elastic properties of 2D materials has also been reported in previous studies, such as 2H MoS$_2$ \cite{cooper2013}, 2H $MX_2$ \cite{li2013}, and 1T$^\prime$ WTe$_2$ \cite{wang2017}. $\sigma_{y}^{\rm max}$ and $\varepsilon_{y}^{\rm max}$ (armchair direction) are larger than $\sigma_{x}^{\rm max}$ and $\varepsilon_{x}^{\rm max}$ (zigzag direction), respectively. The elastic anisotropy is also confirmed in the present work. 

\section{Conclusion}
We have studied the structural properties of 1T$^\prime$-structured $MX_{2(1-x)}Y_{2x}$ ($M=$ Mo and W; $X, Y=$ S, Se, and Te; $0\le x \le 1$) using first-principles approach. We have shown that (i) a universal correlation exists between the formation energy and the distortion amplitude of the 1T$^\prime$ structure and (ii) the site preference of Te atoms plays an important role in stabilizing the 1T$^\prime$ phase. Such preference influences the linear elastic constants, while no such correlations are observed in the non-linear regime. The present work indicates that the structure-elastic property relationship is valid within the linear elastic regime. 

\begin{acknowledgments}
This work was supported by JSPS KAKENHI (Grant No. 24K01142). Calculations were done using the facilities of the Supercomputer Center, the Institute for Solid State Physics, the University of Tokyo and the Supercomputer system at the Information Initiative Center, Hokkaido University, Sapporo, Japan. 
\end{acknowledgments}

\section*{Data Availability}
The data that support the findings of this study are available from Ref.~\cite{zenodo} and the corresponding author upon reasonable request.



\end{document}